\documentclass[11pt]{article}
\usepackage{amsmath,amssymb,amsfonts,epsfig,graphicx,euscript}%,mathrsfs}
\newcommand{\bse}{\begin{subequations}}
\newcommand{\ese}{\end{subequations}}
\newcommand{\be}{\begin{equation}}
\newcommand{\ee}{\end{equation}}
\newcommand{\bea}{\begin{eqnarray}}
\newcommand{\eea}{\end{eqnarray}}
\newcommand{\ba}{\begin{array}}
\newcommand{\ea}{\end{array}}

\newcommand{\ie}{{\it i.e.}}

\newcommand{\eq}[1]{(\ref{#1})}

\begin{document}
\begin{titlepage}
\thispagestyle{empty}

\vspace{2cm}
\begin{center}
\font\titlerm=cmr10 scaled\magstep4 \font\titlei=cmmi10
scaled\magstep4 \font\titleis=cmmi7 scaled\magstep4 {
\Large{\textbf{On Holographic Non-relativistic Schwinger Effect}
\\}}
\vspace{1.5cm} \noindent{{%\large
Kazem Bitaghsir
Fadafan$^{}$\footnote{e-mail:bitaghsir@shahroodut.ac.ir }, Fateme
Saiedi$^{}$\footnote{e-mail:fatimasaiedi@yahoo.com  }
}}\\
\vspace{0.8cm}

{\it ${}$Physics Department, University of Shahrood, Shahrood,
Iran\\}

\vspace*{.4cm}

\end{center}
\vskip 2em

%----------------------------------------------------------------------------
\begin{abstract}
Using the AdS/CFT correspondence, we study the Schwinger effect in
strongly coupled theories with an anisotropic scaling symmetry in
time and spatial direction. We consider Lifshitz and hyperscaling
violation theories and use their gravity duals. It is shown that the
shape of the potential barrier depends on the parameters of theory.
One concludes that the production rate for the pair creation of
particle and anti-particle will not be the same as the relativistic
case.

\end{abstract}

\end{titlepage}
%\tableofcontents

%%%%%%%%%%%%%%%%%%%%%%%%%%%%%%%%%%%%%
\section{Introduction}
%%%%%%%%%%%%%%%%%%%%%%%%%%%%%%%%%%%%%
One of the interesting effects in QED is pair production in a
constant electric field which is known as Schwinger effect
\cite{Schwinger:1951nm}. This is a non-perturbative effect in
quantum field theories. The pair production of charged particles
like particle $q$ and anti-particle $\bar{q}$ with mass $m$ is
exponentially suppressed for
homogeneous background electric filed $E$ as %
\be P \varpropto e^{\frac{-\pi m^2}{|e E|}}\ee%
where $e$ is the charge of the particle. It is clear that there is a
tunneling process with a classical Euclidean action. Increasing the
electric field leads to increasing the production rate so that there
is not a phase transition. There is a smooth behavior even for
larger values of $|e E|$ where the saddle-point approximation ceases
to be applicable. This is not the case in string theory, where there
exists a critical value for the electric field $E_c$
\cite{Fradkin:1985ys,Bachas:1992bh}. This critical value is
proportional to the string tension $|e E_c|\sim \frac{1}{2 \pi
\alpha'}$. As a result, at $E_c$ phase transition occurs and one
finds an instability. It means that the pair production drops to
zero and potential barrier disappears. One should notice that this
instability happens even for a neutral string with opposite charges
at the ends.

Based on $AdS/CFT$ correspondence \cite{CasalderreySolana:2011us},
one may study this instability in some field theories. This study
has been done for $\mathcal{N}=4$ super Yang-Mills (SYM) in four
dimensions in \cite{Semenoff:2011ng}. They consider SYM theory in
the Coulomb branch and spontaneously break the gauge group $SU(N+1)$
to $SU(N)\times U(1)$. Then turn on an electric field of the $U(1)$
gauge theory and pair creation of the massive W-bosons is similar to
the Schwinger effect. It was shown that the critical
value of the electric field is expressed as %
\be E_c=\frac{2 \pi m^2}{\sqrt{\lambda}} \ee%
where $m$ and $\lambda$ are W-boson mass and 't Hooft coupling,
respectively. The $\frac{1}{\sqrt{\lambda}}$ correction to this
value has been studied in \cite{Ambjorn:2011wz}.\footnote{From the
AdS/CFT correspondence, the 't Hoof coupling $\lambda$ is related to
the curvature radius $L$ and the string tension $(\frac{1}{2\pi
\alpha'})$ by $\sqrt{\lambda}=\frac{L^2}{\alpha'}$. }

The Schwinger effect in the context of $AdS/CFT$ correspondence has
been studied in different papers. The universal aspects of this
effect in the general backgrounds are studied in
\cite{Sato:2013hyw}. The pair production in confining geometries is
investigated in \cite{Sato:2013dwa} and \cite{Kawai:2013xya}. They
find two kinds of critical electric field in this case. The
potential barrier for the pair creation is analyzed in
\cite{Sato:2013iua}.  The effect of pair production in the
conductivity of a system of flavor and color branes is discussed in
\cite{Chakrabortty:2014kma}. In de Sitter space time also the
Schwinger effect has been studied in \cite{Fischler:2014ama}. This
effect is also studied in the world-line formalism of quantum field
theory in \cite{Dietrich:2014ala} and in \cite{Gordon:2014aba} as a
WKB exact path integral. The holographic Schwinger effect with
constant electric and magnetic fields has been studied in
\cite{Bolognesi:2012gr} and \cite{Sato:2013pxa}. The vacuum
instability in the presence of a constant electric field in
$\mathcal{N}=2$ supersymmetric QCD has been studied in
\cite{Hashimoto:2013mua}. For study of this effect in the
Sakai-Sugimoto model, see \cite{Hashimoto:2014yya}. They compute the
creation rate of the $q\bar{q}$ pair by evaluating the imaginary
part of the Dirac-Born-Infield (DBI) action including a constant
electromagnetic field. On the other hand, the authors of
\cite{Semenoff:2011ng} evaluate the Nambu-Goto action on the probe
D3-brane.

It is known that methods based on the $AdS/CFT$ correspondence
relate gravity in $AdS_{}$ space to the conformal field theory on
the boundary. These conformal field theories are invariant under the
following scaling transformation%
 \be  (t,\vec{x})\rightarrow (w \,t,w\,\vec{x} ), \ee
where $t$ is time and $\vec{x}$ is spatial directions of the space
time. The scaling factor $w$ is a constant. However, in field
theories near a critical phenomena there is
an anisotropic scaling symmetry as follows:%
 \be (t,\vec{x})\rightarrow
(w^z \,t,w\,\vec{x} ),  \ee%
where $z$ is called dynamical exponent. In the case of $z=1$, theory
shows relativistic scale invariance. %
The non-relativistic geometries have been considered in
\cite{Son:2008ye} and \cite{Balasubramanian:2008dm}. The holographic
description with the Lifshitz fixed point has been studied in
\cite{Kachru:2008yh}. The gravity in 5-dimensions is
described by the following metric%
 \be ds^2=L^2\left(-r^{2z} dt^2 +
\frac{dr^2}{r^2}+r^2d\vec{x}^2\right).
 \label{metricz}\ee%
Where $L$ is the radius of curvature. This geometry has a genuine
null singularity which may be resolved by considering stringy
effects. We ignore these issues and point out that our results may
only be valid in certain range of energies. One should notice that
in the pure cosmological Einstein gravity the space time is isotopic
and to produce an anisotropic space time one should consider other
fields like a massive gauge field. Including both a scalar field
with nontrivial potential and a gauge field lead to the following
metric%
\be ds^2=\frac{L^2}{r^{2\theta/d}}\left(-r^{2z} dt^2 +
\frac{dr^2}{r^2}+r^2d\vec{x}^2\right).
 \label{metrictt}\ee%
where $d$ is the spatial dimension of the boundary and $\theta$ is
hyperscaling violation exponent. To satisfy the null energy
condition, one should assume the following relations between
$\theta$, $z$ and $d$%
\be (z-1)(d+z-\theta)\geq0,\,\,\,\,\,\,(d-\theta)(d\,z-\theta-d)\geq 0.\ee%
Actually, based on the $AdS/CFT$ correspondence non-zero $\theta$
means hyperscaling violation in the dual field theory. An anomalous
scaling dimension is also introduced in
\cite{Karch,Gouteraux:2013oca} to describe the anomalous temperature
scaling of strange metals by studying the frequency and temperature
dependence of the conductivity. Two classes of non-zero anomalous
scaling dimension, are Einstein-Maxwell-Dilaton systems or probe
branes in backgrounds with non-zero $\theta$
\cite{Gouteraux:2012yr}. \footnote{We would like to thank A. Karch
for discussion on this scaling parameter. }

Now we extend the previous studies of the Schwinger effect in
strongly coupled theories with an anisotropic scaling symmetry in
time and spatial direction. We consider Lifshitz and hyperscaling
violation theories and use their gravity duals in eqs. \eq{metricz}
and \eq{metrictt}. We investigate if the potential barrier for
creation of $q\bar{q}$ depends on the dynamical exponent $z$ and
hyperscaling violation $\theta$. Our purpose is to study the shape
of the barrier potential versus parameters of $z$ and $\theta$ to
find how the rate of $q\bar{q}$ changes, qualitatively. We find
analytic equations for the barrier potential and the distance
between $q$ and $\bar{q}$ in these theories. One may even ask
whether the barrier potential disappears by changing $\theta$ and
$z$. We will show that shape of the potential changes and the
production rate for the pair creation will not be the same as the
relativistic case. Interestingly, we find that increasing
hyperscaling parameter $\theta$ and dynamical exponent $z$ have
different effects on the shape of the barrier potential \ie
 by increasing $\theta$ the height and the width of the barrier
increase while by increasing $z$ they decrease, significantly.

This paper is organized as follows. In sections one and two, we will
present the potential analysis in the Lifshitz geometry and
hyperscaling violation, respectively. We study behavior of barrier
potential as a function of distance between $q$ and $\bar{q}$. We
consider the case with $\theta < d$. We also discuss the spacetimes
with anisotropic in spatial direction in section three.
In the last section we summarize our results. %

%%%%%%%%%%%%%%%%%%%%%%%%%%%%%%%%%%%%%%%%%%%%%%%%%%
\section{Potential analysis in Lifshitz geometry}%
%%%%%%%%%%%%%%%%%%%%%%%%%%%%%%%%%%%%%%%%%%%%%%%%%%

To study the holographic Schwinger effect, one may estimate the
barrier potential, holographically.  The general form of this
tunneling barrier depends on the rest masses $2 m$ and the energy
that $q$ and $\bar{q}$ at distance $x$ gained from an external
homogeneous electric field $E$. By adding Coulomb
interactions, the barrier potential has this profile%
\be V=2m -E\,x-\frac{\alpha}{x}\ee%
where $\alpha$ depends on the electric charge. Simple holographic
arguments in \cite{Semenoff:2011ng} leads to an estimate for the
critical electric field which exhibits deviation from the DBI
result. This deviation has been studied by potential analysis in
\cite{Sato:2013iua}. We will follow the same setup and consider a
probe D-brane at finite position in the bulk.

The Schwinger effect in general backgrounds with an external
electric field has been studied in \cite{Sato:2013hyw}. Using
holographic potential analysis, they find some universal aspects for
this effect. Although, we have considered  an anisotropic scaling
symmetry in time and spatial direction but our anisotropic
geometries belong to their class and one expects a critical $E_c$ in
the theory. As it was pointed out, we are going to investigate the
shape of the barrier potential versus parameters of $z$ and
$\theta$. Our purpose is to find how the creation rate of $q\bar{q}$
changes, qualitatively.

Using the $AdS/CFT$ correspondence, we consider $q\bar{q}$ pair on
the D-brane which is located at finite position in the bulk. They
can be realized as the endpoints of the U-shaped string hanging from
the D-brane to the IR-region. The brane covers the boundary
directions and locates at $r_0$ \cite{Sato:2013iua}. We use the
usual orthogonal Wilson lines and assume the $q\bar{q}$ is aligned
in the $x$
direction %
\be t=\tau,\quad x=\sigma,\quad
r=r(\sigma),\label{static-gauge} \ee%
one finds the generic formula for the distance between $q$ and
$\bar{q}$ in \cite{Sato:2013hyw}. Considering the metric in
\eq{metricz}, one obtains the lagrangian from the Euclidean Nambu-Goto action as%
\be \mathcal{L}=L^2\sqrt{r^{2z+2}+r^{2z-2}(\partial_\sigma r)^2}. \ee%
Based on the U-shaped string, there is a turning point at $r_c$
where $\partial_\sigma r=0$. It should be noticed that $r_c<r_0$.
Now one finds the distance $x$ as follows:%

\bea x(z)&=&\frac{2 L^2}{r_c} \int_1^{r_0/r_c}  \frac{dy}{y^2
\sqrt{y^{2z+2}-1}}\nonumber\\&&=\frac{2
L^2}{r_0}[\frac{\sqrt{\pi}\,\Gamma{\left(\frac{2+z}{2+2z}\right)}}
{a\Gamma{\left(\frac{1}{2+2z}\right)}}-\left(\frac{a^{1+z}}{2+z}\right)
\,_2\mathrm{F}_1\left(\frac{1}{2},\frac{2+z}{2+2z},\frac{4+3z}{2+2z},a^{2+2z}\right)].\label{X}\eea%

where $y=\frac{r}{r_c}$ is a new coordinate and $a=\frac{r_c}{r_0}$
measures the tip of the U-shaped string with respect to the location
of the D-brane in the bulk. For the relativistic case of $ z = 1$,
this formula reduces to the case of an isotropic strongly coupled
theory \cite{Sato:2013iua}.

Using the standard calculations of Wilson lines, one finds the sum
of the $q\bar{q}$ static energy and Coulomb potential as follows:\footnote{Higher derivative correction to the $q\bar{q}$ potential has been studied
in \cite{Fadafan:2011gm}}%

\bea V_{1}(z)&=&\frac{r_c^{z}}{\pi \alpha'}\,  \int_1^{r_0/r_c} dy
\frac{ y^{2z}}{\sqrt{y^{2z+2}-1}}\\&&=\frac{r_0^z}{\pi
\alpha'}\,[-\frac{a^z\,\sqrt{\pi}\,\Gamma{\left(\frac{-z}{2+2z}\right)}}
{(1+2z)\Gamma{\left(-1+\frac{1}{2+2z}\right)}}+\left(\frac{1}{z}\right)
\,_2\mathrm{F}_1\left(\frac{1}{2},\frac{-z}{2+2z},\frac{2+z}{2+2z},a^{2+2z}\right)].\label{V1}\nonumber
\eea%
In this potential the UV part of the the string solution is absent,
because we put the D-brane at finite radius in the bulk not in the
boundary. As a result the short distance behavior of $q\bar{q}$ is
modified in this approach. By putting off the D-brane to the
boundary at fixed $r_c$ which means $a \rightarrow 0$, the
Hypergeometric function goes to 1 and the second term in \eq{V1}
reduces to the static energy of infinitely massive $q$ and
$\bar{q}$. The first term will also coincident with the Coulomb
potential in the presence of dynamical $z$ exponent which is
proportional to $1/x^z$. The Wilson loop calculations in the
Lifshitz spacetime which shows the same results for $q\bar{q}$
potential was performed in \cite{Danielsson:2009gi,Kluson:2009vy}.
In the limit $a\rightarrow 1$, Hypergeometric function in the second
term of \eqref{X} goes to
$\frac{\,\sqrt{\pi}(2+z)\,\Gamma{\left(\frac{2+z}{2+2z}\right)}}
{\Gamma{\left(\frac{1}{2+2z}\right)}}$. Then $x$ and $V_1$ vanish.
This is the same as the case of $z=1$ in \cite{Sato:2013iua}.

In the presence of external electric field $E$ along the $x$
direction, the total
potential is given by%

\bea V_{}(z)&=&V_{1}(z)-E\,x(z)\nonumber\\&&=\frac{r_0^z}{\pi
\alpha'}[-\frac{a^z\,\sqrt{\pi}\,\Gamma{\left(\frac{2+z}{2+2z}\right)}}
{z\Gamma{\left(\frac{1}{2+2z}\right)}}+\left(\frac{1}{z}\right)
\,_2\mathrm{F}_1\left(\frac{1}{2},\frac{-z}{2+2z},\frac{2+z}{2+2z},a^{2+2z}\right)\nonumber\\&&
-\frac{b}{a}\left(\frac{\sqrt{\pi}\,\Gamma{\left(\frac{2+z}{2+2z}\right)}}
{\Gamma{\left(\frac{1}{2+2z}\right)}}-\left(\frac{a^{2+z}}{2+z}\right)
\,_2\mathrm{F}_1\left(\frac{1}{2},\frac{2+z}{2+2z},\frac{4+3z}{2+2z},a^{2+2z}\right)
\right)].\label{V1z}\eea

where $b=\frac{E}{E_c}$ and $E_c=\frac{r_0^{z+1}}{2\pi \alpha' L^2}$.%
\begin{figure}[ht]
\centerline{
\includegraphics[width=3.in]{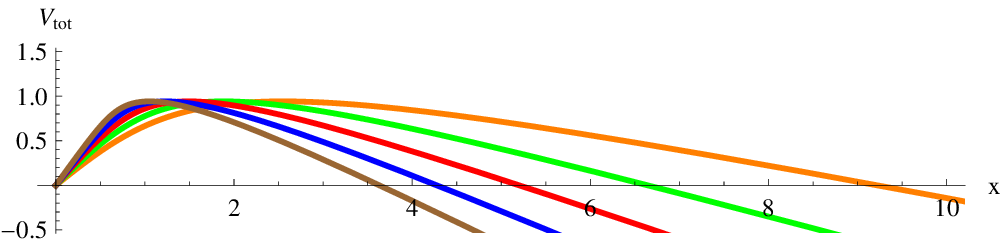}
\includegraphics[width=3.in]{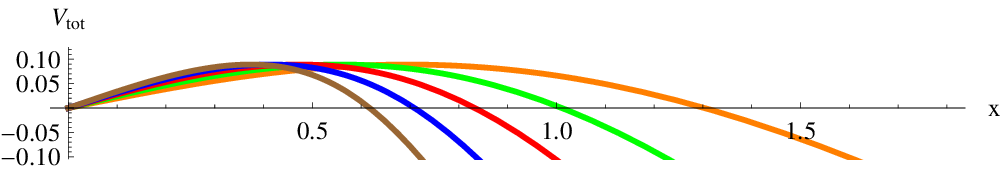}} %
\caption{The barrier potential versus the inter quark distance $x$
in the Lifshitz geometry. Left: $b=0.8$. Right: $b=0.2$. In all of
the plots from right to left $z=1,2,3,4,5$.   }\label{Lifshitz}
\end{figure}%

In Fig. 1, we assume $E<E_c$ and call $V_{tot}$ as barrier potential
in the non-relativistic Schwinger effect. This barrier potential
disappears when $E>E_c$ and vacuum becomes unstable
catastrophically. We plot in this figure the barrier potential
versus the inter quark distance $x$ in the presence of new Lifshitz
scaling symmetry $z$. In the left and right plots, we consider
$b=0.8$ and $b=0.2$, respectively. In all of these plots from top to
bottom the dynamical exponent $z$ increases as $z=1,2,3,4,5$. It is
clear that the height and the width of the barrier depend on the
dynamical exponent $z$. As $z$ increases, the shape of barrier
changes significantly \ie $\,$ the height and the width of barrier
decrease. As a result, one finds that the rate of producing the
$q\bar{q}$ pairs should be changed in the presence of an anisotropic
scaling symmetry in time and spatial direction. We checked that
increasing the electric field, leads to decreasing the height and
the width of the barrier.

Why should one expects that by turning on the dynamical exponent
$z$, the barrier potential also significantly changes? We use some
simple field theory arguments to show that it is expectable. One
should consider that the pair production is a tunneling process
through a barrier of height which is proportional to $2m$. The width
of the barrier is also given in terms of $x\sim\frac{2m}{E}$. The
production rate has exponential suppression which is approximately
$(2m)\,(\frac{2m}{E})\sim \frac{m^2}{E}$. Then the final results
depends strongly on the mass of the quarks.
We see that the mass is given by%
\be m=\frac{\sqrt{\lambda} r_0^z}{2\pi}\ee%
It depends on the dynamical exponent $z$. Then the shape of the
barrier changes by changing $z$.

When the electric field is larger than $E_c$, the barrier potential
disappears. We show the total potential in \eqref{V1z} versus the
distance $x$ in the Fig. 2. As it is clear in this figure, there is
no barrier potential and $q\bar{q}$ pairs create freely and
 the vacuum decays catastrophically. By changing $z$, this phenomena
does not change. Although the shape of the potential depends on $z$.
\begin{figure}[ht]
\centerline{
\includegraphics[width=2.8in]{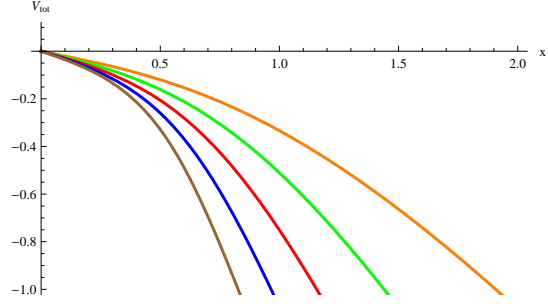}
} %
\caption{The barrier potential versus the inter quark distance $x$
in the Lifshitz geometry for $b=1.2$. From from right to left
$z=1,2,3,4,5$. }
\end{figure}%

%%%%%%%%%%%%%%%%%%%%%%%%%%%%%%%%%%%%%%%%%%%%%%%%%%
\section{Potential analysis in theories with hyperscaling violation}%
\begin{figure}[ht] \centerline{
\includegraphics[width=2.9in]{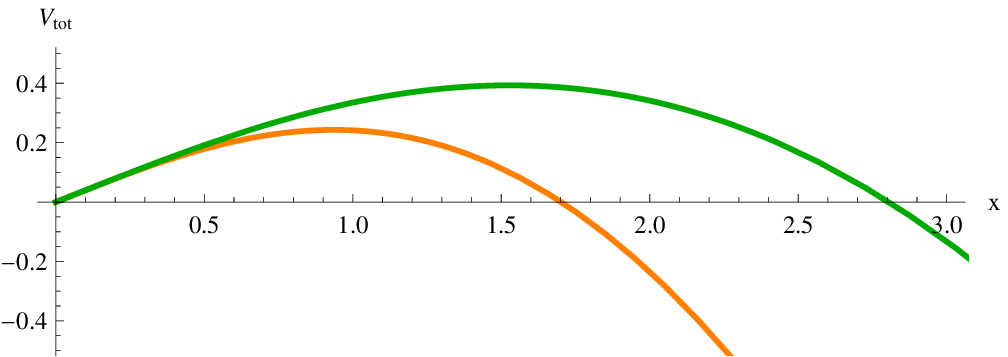}
\includegraphics[width=2.9in]{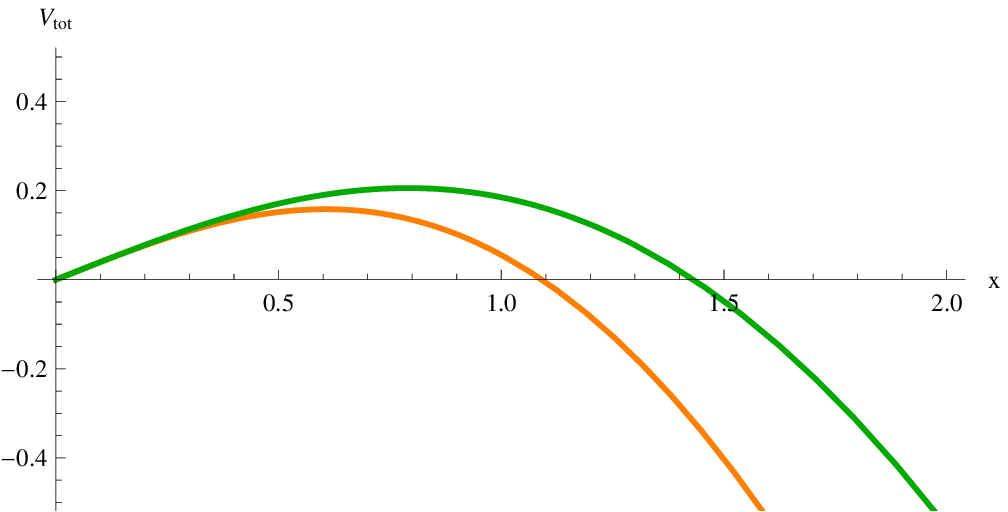}} %
\caption{The barrier potential versus the inter quark distance in
hyperscaling violation geometry. Left: $z=1$. Right: $z=2$. In all
of the plots $d=3$, $b=0.6$ and from top to bottom $\theta=2,1$.
}\label{Hyper}
\end{figure}

%%%%%%%%%%%%%%%%%%%%%%%%%%%%%%%%%%%%%%%%%%%%%%%%%%
Next, we consider barrier potential analysis in hyperscaling
violation geometry in \eqref{metrictt}. We call again the distance
between $q\bar{q}$ pair on the boundary and the $q\bar{q}$ pair
potential energy as $x(\theta,z)$ and $V_{tot}(\theta,z)$,
respectively. One finds the
analytic solutions as follows:%
\bea x(\theta,z)&=&\frac{1}{a} \Gamma\left(\frac{4+2 z-\frac{4
\theta }{d}}{4+4 z-\frac{8 \theta }{d}}\right) \times\\&&
\left(\frac{2 \sqrt{\pi }}{\Gamma\left(\frac{1}{2+2 z-\frac{4 \theta
}{d}}\right)}- \frac{a^{2+z-\frac{2 \theta }{d}} \,}{1+ z-2 \theta/d
}_2\mathrm{F}_1\left(\frac{1}{2},\frac{4+2 z-\frac{4 \theta
}{d}}{4+4 z-\frac{8 \theta }{d}},\frac{8+6 z-\frac{12 \theta
}{d}}{4+4
z-\frac{8 \theta }{d}},a^{2+2 z-\frac{4 \theta }{d}}\right)\right).\nonumber\eea%
and %
\bea &&V_1(\theta , z)=2 a^{z-\frac{2 \theta }{d}}
\times\\&&\left(-\frac{d \sqrt{\pi }
 \Gamma\left(\frac{-d z+2 \theta }{2 (d+d z-2 \theta )}\right)}{(d+2 d z-4 \theta )
  \Gamma\left(\frac{-d-2 d z+4 \theta}{2 d+2 d z-4 \theta }\right)}+\frac{a^{-z+\frac{2 \theta }{d}}
  \,_2\mathrm{F}_1\left(\frac{1}{2},\frac{-d z+2 \theta }{2 (d+d z-2 \theta )},\frac{d (2+z)-2 \theta }{2 (d+d z-2 \theta )},a^{2+2 z-\frac{4 \theta }{d}}\right)}
  { z-2 \theta/d }\right).\nonumber\eea%
where $V_{tot}(\theta,z)=V_1(\theta,z)-E\,x(\theta,z)$. In
Fig.\ref{Hyper}, we show the effect of the hyperscaling parameter
$\theta$ on the barrier potential. In the left and right plots, the
dynamical exponent is $z=1$ and $z=2$, respectively. It is clearly
seen that by increasing $z$ the hight and width of the barrier
decreases. One finds also that increasing $\theta$ leads to
increasing the hight and the width of the barrier. It is interesting
that increasing hyperscaling parameter $\theta$ and dynamical
exponent $z$  have different effects on the shape of the barrier
potential. Then one can change the shape of the barrier potential by
changing the values of these parameters. As a result, the rate of
producing the $q\bar{q}$ pairs depends on these parameters. It means
that pair production is easier when $z$ increases.

%%%%%%%%%%%%%%%%%%%%%%%%%%%%%%%%%%%%%%%%%%%%%%%%%%%%%%%%
\section{Anisotropic scaling in spatial direction}%%%%
%%%%%%%%%%%%%%%%%%%%%%%%%%%%%%%%%%%%%%%%%%%%%%%%%%%%%%%%
We have considered an anisotropic scaling symmetry in time and
spatial direction. It would be interesting to discuss the case of
anisotropic scaling in one of the spatial directions. Then one may
follow \cite{deBoer:2011wk,Alishahiha:2012cm} and consider
anisotropic scaling in one of the spatial directions. Using a double
Wick rotation in \eqref{metrictt} as $t \rightarrow iy$ and
$x_d\rightarrow it$ one finds the corresponding background as%
\be
ds^2=\frac{L^2}{r^{2\theta/d}}\left(-r^2\,dt^2+\frac{dy^2}{r^{2z}}
+r^2d\vec{x}^2+ \frac{dr^2}{r^2}\right).
 \label{metrictat}\ee%
We turn on the external electric field in $x$ and $y$ directions,
respectively. It is straightforward to find the related equations in
this geometry. If we assume $q\bar{q}$ pair in the $x$ direction and
turn on the external electric field, one finds the same behavior in
the Fig. \ref{Lifshitz}. It means that by increasing $\theta$ the
barrier potential becomes stronger. Next, we consider the $q\bar{q}$
pair in the $y$ direction and turn on the electric field. In this
case both of parameters \ie $\,\theta$ and $z$ appear in the total
potential equation. Then by changing them the shape of the barrier
also changes, significantly.

The type IIB solutions dual to Lifshitz theories with anisotropic
scaling symmetry have been discussed in \cite{Azeyanagi:2009pr}.
They consider the following anisotropic spacetime%
\be ds^2=-r^{2z}dt^2+r^{2z}d\vec{x}^2+r^2 dy^2+\frac{dr^2}{r^2}.\ee%
These solutions describe anisotropic theory in the IR regime and the
$AdS_5$ solutions in the UV regime. They considered geometries with
intersecting D3 and D7 branes. The energy loss of heavy quarks in
this background has been studied in \cite{Fadafan:2009an}. The
computation of the total potential of the Schwinger effect in this
background is straightforward and one finds the same results as
before. Simply, one should turn on the electric field in the
different directions and study whether the barrier potential gets
stronger or weaker.

%%%%%%%%%%%%%%%%%%%%%%%%%%%%%%%%%%%%%%%%%%%%%%%%%%%%%%%%%%%%%%%%%%
\section{Conclusion}
%%%%%%%%%%%%%%%%%%%%%%%%%%%%%%%%%%%%%%%%%%%%%%%%%%%%%%%%%%%%%%%%%%
In this paper, we have studied the holographic Schwinger effect in
the non-relativistic setting. We considered Lifshitz and
hyperscaling violation theories. They are strongly coupled theories
with an anisotropic scaling symmetry in time and spatial direction.
An understanding of how this effect changes by these theories may be
essential for theoretical predictions. We discussed how the barrier
potential changes in the presence of non-relativistic parameters $z$
and $\theta$. One motivation for studying the behavior of the
barrier potential is that the tunneling trajectory which is a circle
in the relativistic case, may not be so simple any more for $z\neq
1$. We found the analytical solutions for distance $d$ and total
potential. It was shown that the shape of the potential barrier
depends on these parameters. As a result, we found qualitatively how
the production rate of $q\bar{q}$ pairs change. In this way, one
concludes that the creation rate for the pair creation will not be
the same as the relativistic case. By increasing $z$, $q\bar{q}$
pairs produce easier while increasing $\theta$ leads to harder
production of $q\bar{q}$ pairs. It will be very interesting to
investigate the creation rate of the $q\bar{q}$ pair by evaluating
the imaginary part of the DBI action in the case of the Lifshitz
geometry. Also the analysis of the barrier potential could be
generalized to the finite temperature.
\section*{Acknowledgment}
The authors thank M. Ali-akbari, M. Atashi, K. Karch, Y. Sato, A.
Sonoda, K. Yoshida and K. Zarembo for discussions on different
aspects of holographic Schwinger effect.
%%%%%%%%%%%%%%%%%%%%%%%%%%%%%%%%%%%%%%%%%%%%%%%%%%%%%%%%%%%%%%%%%%

%%%%%%%%%%%%%%%%%%%%%%%%%%%%%%%%%%
\end{document}